\documentclass{osa-article}

\journal{osac}

\articletype{Research Article}

\begin{document}

\title{Shaping entangled photons through emulated turbulent atmosphere}

\author{Ronen Shekel,\authormark{1} Ohad Lib,\authormark{1} Alon Sardas,\authormark{1} and Yaron Bromberg\authormark{1,*}}

\address{\authormark{1}Racah Institute of Physics, The Hebrew University of Jerusalem, Jerusalem, 91904 Israel}

\email{\authormark{*}yaron.bromberg@mail.huji.ac.il} 



\begin{abstract}
Scattering by atmospheric turbulence is one of the main challenges in creating long free-space optical links, and specifically links of entangled photons. Classical compensation methods are hard to apply to entangled photons, due to inherently low signal to noise ratios and the fragility of entanglement. We have recently shown that we can use the bright laser beam that pumps spontaneous parametric down conversion to control the spatial correlations between entangled photons for compensating their scattering. In this work, we apply the pump-shaping technique to compensate for scrambling of correlations between entangled photons that scatter by emulated atmospheric turbulence. We use a spatial light modulator and Kolmogorov's turbulence model to emulate atmospheric turbulence in the lab, and enhance the entangled photons' signal by a factor of fifteen using pump optimization. We show this for both static and dynamic emulated atmosphere, and demonstrate also the compensation of the scattering of a higher-order mode. Our results can open the door towards realizing free-space quantum links with entangled photons, used in applications such as quantum key distribution.  
\end{abstract}

\section{Introduction}
Quantum technologies hold great promise for revolutionizing a variety of technologies \cite{Dowling2003,OBrien2009}. Quantum computers are expected to disrupt some of the main paradigms of cryptography \cite{bernstein2009introduction}, while quantum encrypted links just might be the next generation of secure communications, implementing quantum key distribution (QKD) protocols \cite{bennett1984quantum,shor2000simple}.

The technical primitive needed for many QKD protocols is to create a communication link of single or entangled photons. Free-space links can potentially exhibit much lower loss than optical fibers\cite{chen2020sending}, thus allowing communication over longer distances, as has been shown experimentally in a satellite-to-ground constellation \cite{Liao2017}. In particular, the use of the spatial degree of freedom is attractive, due to its high dimensionality, and its relative ease of control\cite{walborn2006,erhard2020advances,leed2020programmable,lib2020spatially,brandt2020high,boucher2021spatial}. 

One of the challenges in building a free-space communication link is atmospheric turbulence, which causes both spatial and temporal variations in the refractive index of the atmosphere. For classical coherent light, the turbulence leads to scattering, resulting in aberrations. When the scattering is significant a speckle pattern is observed. In the case of entangled photons the spatial correlations encoding the information are distorted, yielding a two-photon speckle pattern \cite{peeters2010observation,gnatiessoro2019imaging,lib2020thermal}. 

While quantum free-space communication is in its infancy, classical free-space optical communication is well developed. Specifically, adaptive optics \cite{tyson2015principles} and wavefront shaping \cite{Vellekoop2007,mosk2012controlling} were developed to overcome the effects of atmospheric turbulence.  While the adoption of classical wavefront shaping techniques to the quantum domain is not trivial, due to the inherently low signal to noise ratio of quantum signals, over the past few years significant progress in compensating for the scattering of single\cite{Defienne2014,Wolterink2016,Defienne2016,liu2019single} and entangled \cite{defienne2018adaptive,lib2020real,lib2020pump,valencia2020unscrambling} photons has been made. In particular, we have recently demonstrated real-time shaping of entangled photons scattered by a dynamic diffuser\cite{lib2020real}, by shaping the bright pump beam that stimulates the generation of entangled photon pairs in spontaneous parametric down conversion. In this work, we apply the pump-shaping approach to compensate for scattering induced by emulated atmospheric turbulence. We use Kolmogorov and von-Karman's model to emulate atmospheric turbulence with a spatial light modulator (SLM), and show that pump shaping exhibits an enhancement factor of fifteen in the correlations measured by entangled photons scattered by the SLM. We also demonstrate the compensation of the scattering of a higher-order mode. Finally, we demonstrate that the optimization process works also in the case of a dynamic, moving atmosphere. 

\section{Pump shaping}
We begin by explaining the quantity being measured in the photon pair case. A pump beam propagates through a non-linear crystal, where a spontaneous parametric down conversion (SPDC) process occurs, and a pair of spatially entangled photons with double the wavelength are created. The correlations are thus measured using two single photon detectors. The first detector is kept stationary, and the second detector scans an area around the location where the matching correlated photon should be. The coincidence rate where photons are found in both locations simultaneously is then counted. The scattering caused by the turbulent atmosphere ruins these correlations, and measuring them produces a so-called two-photon speckle pattern \cite{peeters2010observation}.

For classical light, scattering can be compensated for by using wavefront shaping with an SLM \cite{Vellekoop2007}. For example, using a feedback algorithm, the phases of the SLM pixels can be optimized to maximize the power at a desired focal spot \cite{Vellekoop2008}. 

\begin{figure}[htbp]
\centering\includegraphics[width=12cm]{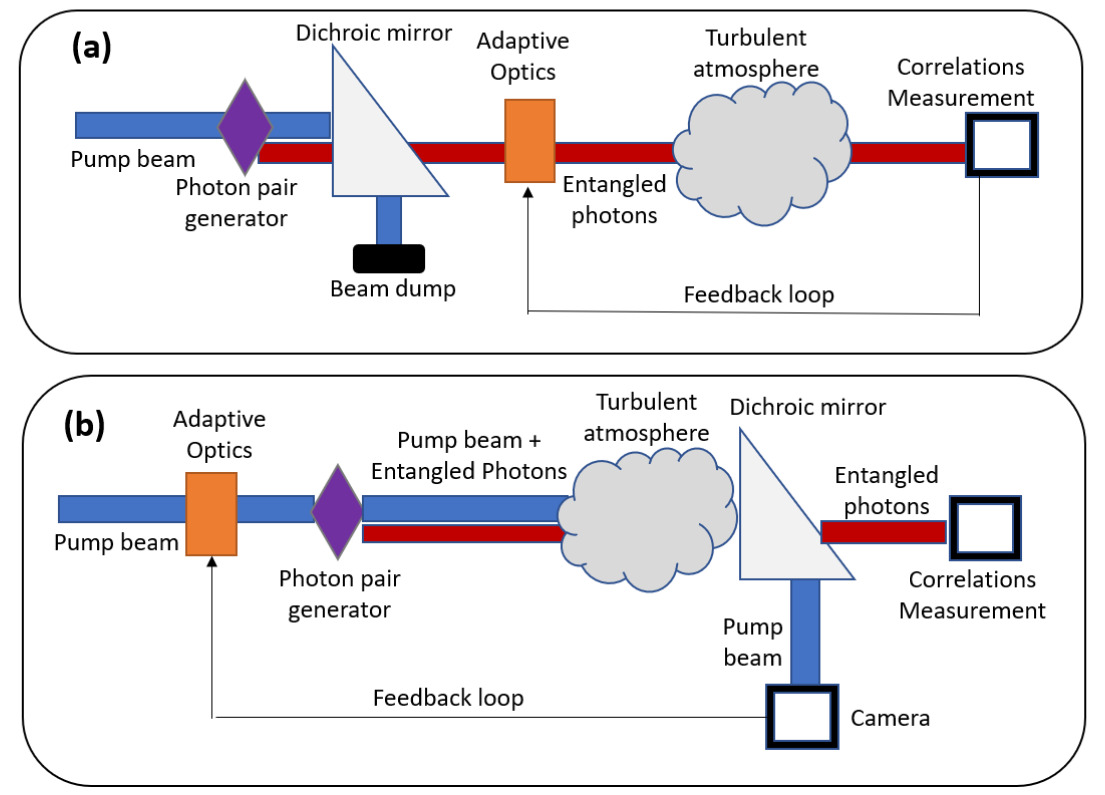}
\caption{\label{fig:idea}(a) Schematic configuration to compensate for scattering of entangled photons induced by atmospheric turbulence. Wavefront shaping is applied directly on the photon pairs, and the feedback signal is provided by the correlations measurement between the photons. (b) Using the new pump-shaping approach, the wavefront shaping is applied to the bright pump beam (blue) which propagates through the same random media as the entangled photons (red). The dichroic mirror is positioned at the target to split the beams. An optimization algorithm is employed using the pump signal, and as a result also the entangled photons correlations are enhanced.}
\end{figure}

In principle, for quantum light one could use the exact same technique and feedback algorithm, measuring the strength of the correlations instead of the power of the beam at the focus, as depicted in Fig. \ref{fig:idea}a. However, since the signals associated with light at the single photon regime are inherently low, in practice this approach is impractical. To address this challenge, we have recently developed the pump-shaping method to compensate for scattering of entangled photons \cite{lib2020real,lib2020pump}. As depicted in Fig. \ref{fig:idea}b, in pump shaping we send the pump beam that stimulates generation of entangled photon pairs via SPDC, together with the entangled photons through the scattering media. We then apply wavefront shaping to focus the bright pump beam, and this simultaneously localizes the quantum two-photon correlations. 

The success of this method is due to two effects. First, the shaping done on the pump beam must be transferred to the correlations between the entangled photons. This is a well known phenomenon in SPDC in the thin crystal regime, leading to applications such as quantum state engineering \cite{molina2005control, valencia2007shaping, kovlakov2018quantum, defienne2019spatially, lib2020spatially, lib2020pump, lib2020real, boucher2021spatial}, controlling orbital angular momentum entanglement \cite{Monken1998, romero2012orbital} and simulating correlations of structured photon pairs \cite{zhang2018boydpumpcoher, trajtenberg2020simulating}. The second effect is that there is a correspondence between the scattering of the pump beam and the entangled photons, as was shown in our recent works \cite{lib2020pump, lib2020real, lib2020spatially}. 

Pump shaping therefore enables optimizing the quantum signal by controlling and measuring only the classical pump beam, offering two main advantages over direct shaping of the photon pairs: i. The feedback signal is many orders of magnitude stronger. ii. The wavefront shaping apparatus induces the loss to the classical bright pump beam and not to the weak flux of photon pairs. 

\section{Emulated atmospheric turbulence}
The effect of atmospheric turbulence on the propagation of the entangled photons can be emulated using a single phase-only mask. While not demonstrating effects such as the finite isoplanatic patch of a real atmospheric link, a single-screen model can faithfully reproduce the statistics of the light scattered by the turbulent atmosphere\cite{burger2008simulating,malik2012influence,ren2014adaptive}, enabling us to study the effect of turbulence using a single SLM, as depicted in Fig. \ref{fig:setup}a. 

The phase distortions to be displayed on the SLM are computed using Kolmogorov and von-Karman's turbulence model\cite{goodman2015statistical, Rickenstorff:16}. Kolmogorov's model predicts the fluctuation in the refractive index of the atmosphere, which occur mainly due to temperature fluctuations, and are described as turbulent eddies. Within the inner scale of the turbulence the power spectrum density of the refractive index fluctuations is described by $\Phi_n(k)=0.033C_{n}^{2}k^{-11/3}$. For values of $k$ below a critical value $k_o$, equivalent to very large size scale $l_o=2\pi/k_o$, the model does not predict the shape of $\Phi_n$, since it becomes highly dependant on large scale geographic and meteorological conditions. Conversely, for $k$ values larger than a critical value $k_m$, equivalent to a scale size $l_i\approx5.92/k_m$ the turbulent eddies dissipate their energy due to viscous forces. To include the effects of these inner and outer scales, the \textit{Von Karman spectrum} is often adopted: $\Phi_{n}(k)=\frac{0.033C_{n}^{2}}{\left(k^{2}+k_{o}^{2}\right)^{11/6}}\exp\left(-\frac{k^{2}}{k_{m}^{2}}\right)$.

To simulate the propagation of photons through atmospheric turbulence, the three-dimensional refractive index fluctuations can be transferred into a two-dimensional phase screen with a PSD of the form\cite{Rickenstorff:16}:

\begin{equation}\label{eq:kolmogorov PSD}
\Phi\left(k_{x},k_{y}\right)=0.49r_{0}^{-\frac{5}{3}}\left(k_{x}^{2}+k_{y}^{2}+k_{o}^{2}\right)^{-\frac{11}{6}}\exp\left(-\frac{k_{x}^{2}+k_{y}^{2}}{k_{m}^{2}}\right)
\end{equation}

where $r_{0}=\left(0.4229\left(2\pi/\lambda\right)^{2}zC_{n}^{2}\right)^{-3/5}$ is the atmosphere coherence width, often called the Fried parameter. $C_{n}^{2}$ is the structure constant of the refractive index fluctuations, and $z$ is the length of the atmospheric link. The outer scale $l_{o}$ is typically in the 1-100m range, and the typical value of the inner scale $l_{i}$ is a few millimeters \cite{goodman2015statistical}. 

To compute phase patterns that yield fields which follow the PSD described by Eq. \ref{eq:kolmogorov PSD}, we use the inverse Fourier transform method. In this method, the square root of the power spectral density distribution given in Eq. \ref{eq:kolmogorov PSD} is multiplied element-wise by a circular Gaussian random matrix, i.e. a matrix of the form $A+iB$ where $A$ and $B$ are real matrices with random normally distributed elements. The real (or imaginary) part of the inverse Fourier transform of the resulted matrix is the desired phase screen for emulating the effect of turbulence \cite{lane1992simulation}. An example of a phase mask that yields a speckle pattern with a PSD that follows Eq. \ref{eq:kolmogorov PSD} is depicted in Fig. \ref{fig:setup}b.

\section{Results}
\subsection{Experimental setup}
\begin{figure}[htbp]
\centering\includegraphics[width=\linewidth]{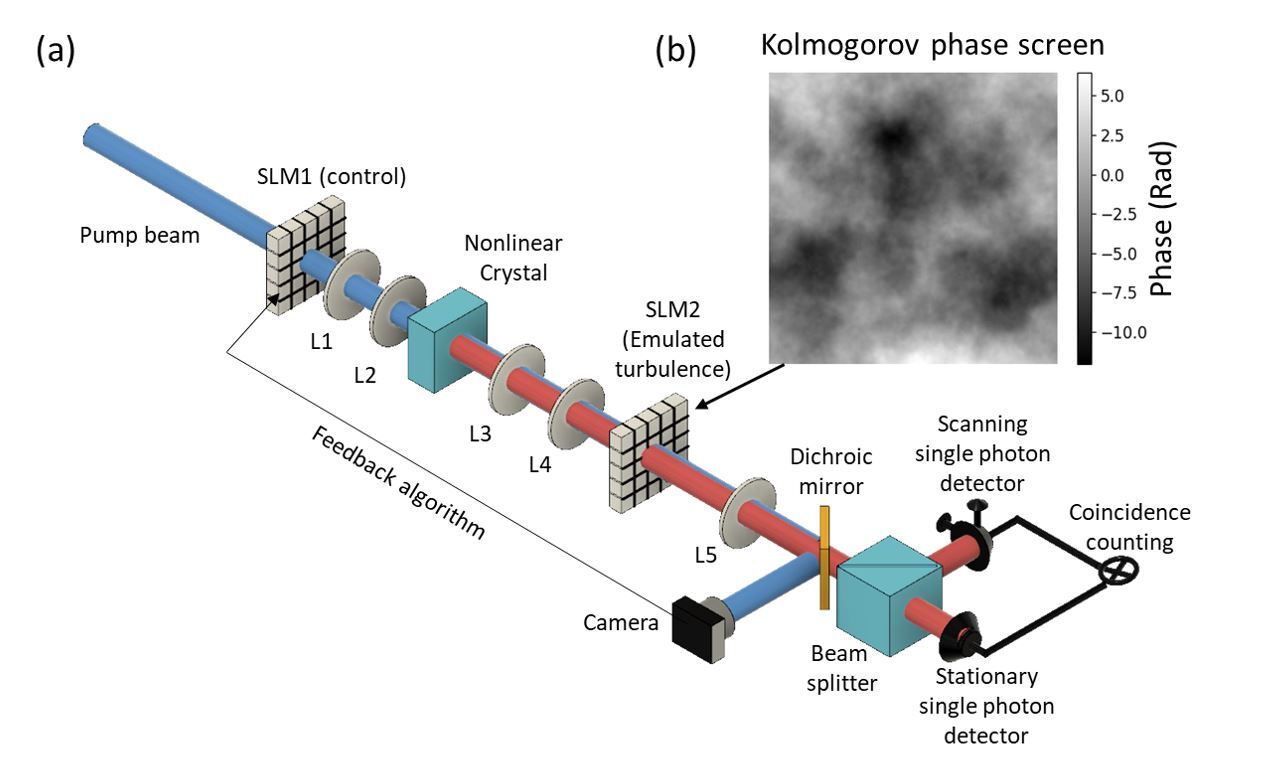}
\caption{\label{fig:setup}(a) Experimental setup. Spatially entangled photons are created by pumping a nonlinear crystal (PPKTP) with a $\lambda=404$nm continuous-wave laser via spontaneous parametric down conversion. Both the pump beam and the entangled photons pass through SLM2 that emulates the atmosphere, which is imaged on the crystal and control SLM1 planes by lenses L1-L4, and measured at the far-field (lens L5). (b) A sample phase screen producing Kolmogorov PSD statistics, with $C_{n}^{2}=10^{-15}$, $z=1$km, $l_o=10$m and $l_i=5$mm.}
\end{figure}

The experimental setup is based on the setup presented in \cite{lib2020real} where we have replaced the scattering diffuser with an SLM that emulates the atmospheric turbulence, as is depicted in Fig. \ref{fig:setup}. Specifically, a $2~\mathrm{mm}$ long type-0 PPKTP crystal is pumped by a $14 ~\mathrm{mW}$, $\lambda=404~\mathrm{nm}$ (blue) continuous-wave laser, generating spatially entangled photons via SPDC (with a Schmidt number of $\approx 680$, as measured in \cite{lib2020real}). The wavefront of the pump beam is shaped by a phase-only SLM (SLM1), imaged on the crystal by two lenses with focal lengths $L1=200~\mathrm{mm}$ and $L2=100~\mathrm{mm}$ respectively. Without shaping, the pump profile at the crystal plane is approximately Gaussian with a waist of $0.7~\mathrm{mm}$. Both the pump beam and the entangled photons are then imaged onto another SLM (SLM2) which emulates the atmospheric turbulence, by two lenses with focal lengths $L3=100~\mathrm{mm}$ and $L4=200~\mathrm{mm}$. Both SLMs feature a pixel size of $12.5 \mu m$, and the beam waist on them is $\approx1.4~\mathrm{mm}$. 

The pump beam and the entangled photons are separated using a dichroic mirror and measured at the far-field by a CMOS camera (pixel size of $4.8 \mu m$) and $50~\mathrm{\mu m}$ multimode fibers coupled to single photon detectors, respectively. The far-field measurements are obtained after passing through a $L5=300~\mathrm{mm}$ lens. For the coincidence measurements, $10~\mathrm{nm}$ interference filters around $808~\mathrm{nm}$ are used, and the coincidence window is $4~\mathrm{ns}$. Both measurements are made off-axis by adding a linear phase to the desired phase mask to discard specular reflection from SLM2 due to its finite diffraction efficiency.

The crystal was chosen to be type-0 so that the polarization of the entangled photons and the pump beam is identical, allowing both of them to be affected by SLM2 without any further polarization manipulations. However, for real atmospheric links, polarization effects are negligible \cite{boyer1978atmospheric}, so type-1 and type-2 processes can be used as well. 

The length of the crystal is chosen as a compromise between the obtained signal levels and the necessity to be within the thin crystal regime in which the angular spectrum of the pump beam is transferred to the entangled photons \cite{Monken1998}. Since the crystal is $1\mathrm{mm}$ wide and $2\mathrm{mm}$ tall it clips the beam, resulting in the elliptical shape seen in the far-field. 

The ratio between the beam size and the pixel size of the camera and SLMs was chosen to cover enough pixels to attain good resolution. On the SLMs $\approx 40000$ pixels manipulate the beam, and the pump speckle grain is resolved by the CMOS camera using $\approx 100$ pixels. When scanning the SPDC signal we use a step size of $25 \mathrm{\mu m}$, so the two-photon speckle is resolved using $\approx 25$ pixels.

The speed of the optimization algorithm is limited by the response time of the liquid crystals in the SLM and by the control electronics. In the dynamic case, also the integration time for measuring the SPDC signal with sufficient signal to noise ratio limits the optimization speed.

\subsection{Speckle measurements}
\begin{figure}[htbp]
\centering\includegraphics[width=\linewidth]{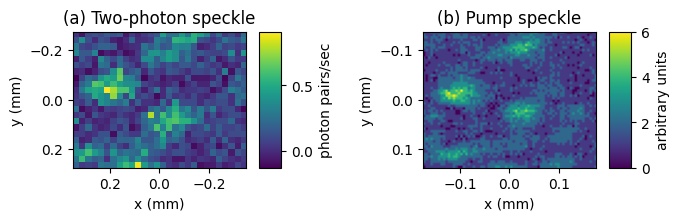}
\caption{\label{fig:speckles}The coincidence pattern (a) and far-field pump beam intensity (b) after passing through SLM2 emulating atmospheric turbulence (with $C_{n}^{2}=10^{-15}$ and $z=1$km). The signals are clearly correlated, up to a spatial scaling factor of two, originating from their wavelength difference. The exposure time of the SPDC signal is $30$ sec/pixel, and of the pump beam is $200~\mathrm{\mu s}$ per frame.}
\end{figure}

We start by comparing the two-photon speckle obtained by scattering by the phase screen imprinted on SLM2 and the classical speckle pattern obtained by scattering of the pump beam by the same phase screen. As seen in Fig. \ref{fig:speckles}, the two-photon speckle is clearly correlated to the pump speckle, up to a spatial scaling factor of two, which results from the fact that the wavelength of the signal and idler photons is two times longer than the pump wavelength. 

\subsection{Optimization for static atmosphere}
Next, we apply an optimization algorithm, using SLM1 to fix the scattering. We emulate a turbulent 1km link with a structure constant $C_{n}^{2}=10^{-15}$, and use $30\times30$ pixels on the control SLM1 to employ the partitioning optimization method \cite{Vellekoop2008}. In each iteration of the algorithm we randomly choose half of the pixels on the SLM, and change their phase relative to the other half. We sample the intensity at the target area for the different relative phases, and estimate the optimal phase based on the expected sinusoidal variation of intensity. This process is repeated until the desired enhancement of the intensity is achieved.

\begin{figure}[htbp]
\centering\includegraphics[width=\linewidth]{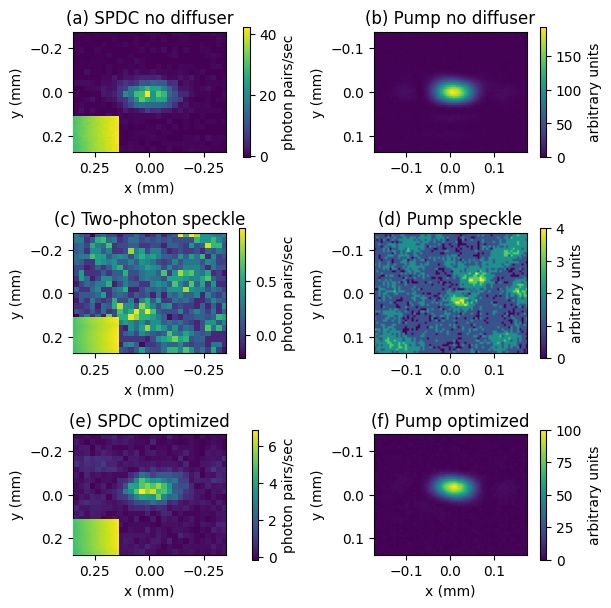}
\caption{\label{fig:optimization}Optimization of quantum correlations. Left column shows the coincidence pattern with no random phase screen (a), after passing through a Kolmogorov phase screen with $C_{n}^{2}=10^{-15}$ and $z=1$km (c), and after the optimization algorithm (e). Right column show the pump beam with no random phase screen (b), after passing through the same phase screen (d), and after the optimization algorithm (f). The exposure times for (a) is $2$ sec/pixel, for (c) $12$ sec/pixel, and for (e) $10$ sec/pixel. All pump images were taken with exposure time of $200~\mathrm{\mu s}$ per frame. Insets are added to the SPDC measurements, showing the single counts in the scanned region. The counts vary between $4500$ and $6500$ counts per second in (a) and (c), and between $3700$ and $5300$ in (e), due to loss caused by the shaping in SLM1 before entering the crystal.}
\end{figure}

The results are shown in Fig. \ref{fig:optimization}. As a reference, we first show the two-photon correlations (Fig \ref{fig:optimization}a) and the pump intensity (Fig \ref{fig:optimization}b) without going through any diffuser. Upon passing through the emulated atmosphere, both the two-photon correlations (Fig. \ref{fig:optimization}c) and the pump intensity (Fig. \ref{fig:optimization}d) exhibit speckle patterns. After applying the optimization algorithm, we achieve a clear enhancement in the correlation signals (Fig. \ref{fig:optimization}e), and see that the pump beam is also refocused (Fig. \ref{fig:optimization}f). 

The single counts are shown as insets in (a), (c), and (e) of Fig. \ref{fig:optimization}, and as expected \cite{lib2020real} they remain the same throughout the optimization process. 

To assess the affect of the optimization process we introduce the enhancement parameter $\eta$. The enhancement is defined as the signal at the target area after the optimization divided by the average signal before the optimization.

We define our target area using the correlations obtained without a scattering phase mask, by choosing the pixels with counts above one quarter of the maximal pixel. For the power before optimization we calculate the average counts per pixel in the two-photon speckle pattern, and multiply by the amount of pixels in the target area. The power after optimization is calculated by simply summing the counts of the relevant pixels at the target area. 

We also define the efficiency of the optimization process as the total coincidence counts in the relevant pixels after optimization compared to the total coincidence counts in the case without displaying a scattering phase mask at all.

The enhancement of the pump beam in Fig. \ref{fig:optimization} is calculated to be $~\eta=31$, with efficiency of $~40\%$. The enhancement of the SPDC signal in Fig. \ref{fig:optimization} is calculated to be $\eta\approx15$, with efficiency of $\approx15\%$. We note that in all SPDC figures and calculations, we subtract the accidental counts, which are $\approx0.14$ counts per second on average.

We also note that in our experiment the beam intensity waist propagating through SLM2 is $\approx1$mm. To ensure significant scattering of the photons, the parameters of the emulated turbulent atmosphere ($l_{o}$,$l_{i}$,$r_{0}$) were scaled down so that our $1$mm beam scatters equivalently to the extreme case of a large $1$m beam propagating through a real atmospheric link\cite{sun2018active,wang2019laser}. The parameters after scaling are given by $r_0=0.14~\mathrm{mm}$, $l_o=10~\mathrm{mm}$ and $l_i=5~\mathrm{\mu m}$.

\subsection{Creating higher order modes}
Encoding information in the spatial regime leads to higher dimensional qubits. Having large dimensionality is beneficial for the purpose of boosting the capacity of quantum systems, since multi-level quantum bits can carry more information than a single two-dimensional qubit \cite{bechmann2000largeralphabets, cerf2002dlevel, erhard2018twisted, forbes2021structured}. Additionally, high-dimensional qubits present enhanced resilience to noise, as was experimentally shown using the orbital angular momentum basis, and energy-time entanglement  \cite{ecker2019overcoming, Mirhosseini2015}.

\begin{figure}[htbp]
\centering\includegraphics[width=\linewidth]{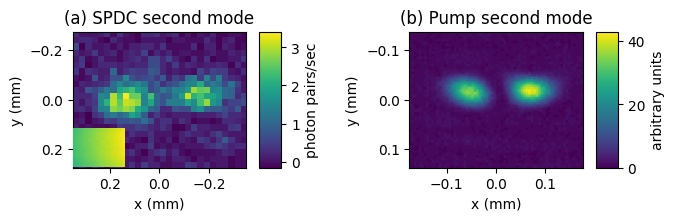}
\caption{\label{fig:second-mode}Shaping the beam post-optimization to attain a higher mode in the coincidence pattern (a) and in the pump beam (b). The exposure time of the SPDC signal is $10$ sec/pixel, and of the pump beam is $200~\mathrm{\mu s}$ per frame. The inset in (a) shows the single counts in the scanned region, where the counts per second vary between $3700$ and $5300$.}
\end{figure}

One of the remarkable features of shaping the pump beam is the ability to encode these higher order spatial modes in the entangled photons \cite{erhard2018twisted,erhard2020advances}. Thus, SLM1 can be used both to encode high order bits of information and to compensate for the atmospheric turbulence, simultaneously. For example, by adding a $\pi$-step to the compensating phase, as depicted in Fig. \ref{fig:second-mode}. The same method can be used in the future for the simultaneous compensation of scattering and diffraction, by combining wavefront shaping and the generation of non-diffracting entangled Airy photons\cite{lib2020spatially}.

\subsection{Optimization for dynamic atmosphere}
The typical way to emulate the dynamics of turbulence using phase screens is based on Taylor's frozen-turbulence hypothesis \cite{schmidt2010numerical}. Under this hypothesis, the shape of the turbulent atmosphere is assumed to be nearly constant (frozen), yet to move in the transverse direction at a speed related to the Greenwood frequency \cite{tyler1994bandwidth}. In our experiment, we create a large Kolmogorov phase screen, and move it transversely with a constant speed (the phase mask is discretely moved by a single pixel every 150 seconds). 

\begin{figure}[htbp]
\centering\includegraphics[width=\linewidth]{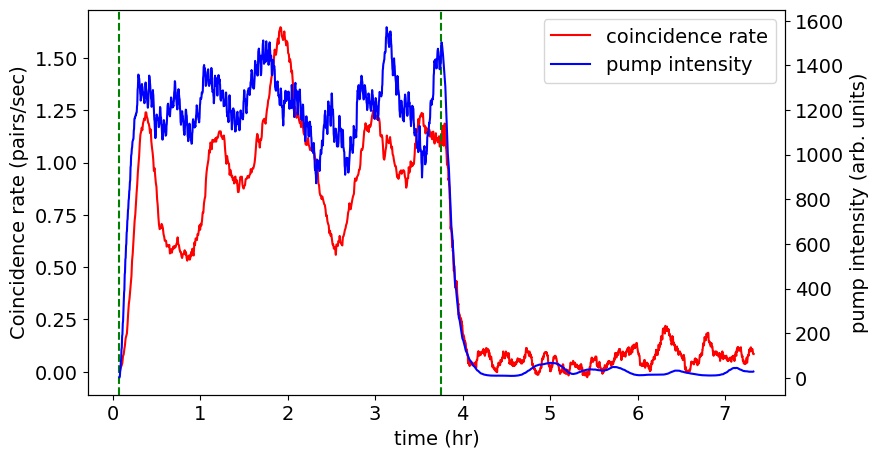}
\caption{\label{fig:moving atmosphere}Correction for an emulated dynamic atmosphere. We apply a moving Kolmogorov phase mask (with $C_{n}^{2}=10^{-15}$, $z=1$km) on SLM2, moving a single pixel every 150 seconds. We apply the pump optimization algorithm (between the vertical green lines) and measure the pump intensity (blue) and the coincidence rate (red). It is clear that the optimization increases both the pump and the two-photon signals.}
\end{figure}

As can be seen in Fig. \ref{fig:moving atmosphere}, once the optimization algorithm is turned on (first green vertical line), both the pump intensity (blue) and coincidence rate (red) rise dramatically. Once the optimization is turned off (second green vertical line), both pump intensity and coincidence rate drop. It can be seen that the duration of the drop is $\approx20~\mathrm{min}$, which is equivalent to a movement of the phase screen by $\approx 0.1\mathrm{mm}$, which is in the order of $r_0$, as expected.

The slow moving speed was chosen due to the limited speed of our optimization and the desire to simultaneously measure the coincidence rate with reasonable SNR. In our current setup the response time of the SLM limits the rate of phase changes to $\approx10Hz$, where many phase changes are needed to achieve optimization. The SPDC signal is on the order of $1$ count per second, so to achieve reasonable SNR we need an exposure time of at least several seconds. Considering these factors, we could not perform the optimization for real atmospheric timescales which are typically in the order of milliseconds \cite{davis1996measurement}. Applying dynamic optimization with a slightly faster moving phase screen will result in a lower enhancement factor, due to less optimization iterations per pixel movement. 

Nevertheless, using fast modulators such as deformable mirrors, digital micromirror devices (DMDs) \cite{conkey2012highDMD} and microelectromechanical systems (MEMS), pump shaping can be run orders of magnitude faster\cite{tzang2019wavefront}, and in principal be scaled up to the relevant rates without any further modifications. 

In this experiment we measure coincidence counts on a single pixel. Therefore, to calculate the enhancement $\eta$ we divide the average power during the time the optimization is working by the average power after the optimization is frozen. We calculate $\eta\approx15$ for the SPDC signal, while the pump enhancement is $\eta\approx45$.

We note that the average counts after stopping the optimization in the dynamic case is lower than the average counts in the static speckle pattern. This is due to the frozen "fixing" phase on SLM1 that loses its correlation with the diffusing phase mask and thus functions as an additional diffusing mask. This is also why the enhancement of $15$ does not bring the signal to the same count value as in the static case. 

\section{Discussion}
We have demonstrated the compensation of the scattering of spatially entangled photons by emulated turbulent atmosphere using the recently proposed pump-shaping approach\cite{lib2020real,lib2020pump}. The compensation was also performed in real-time through dynamic scattering emulated by a moving phase screen on the SLM, and for an higher-order mode. 

The imperfect efficiency of the pump beam optimization can be explained by several experimental imperfections. Mainly, as only $30\times 30$ degrees of freedom are used in SLM1 for compensating the scattering, we expect (according to numerical simulations) a maximal efficiency of $\approx 55\%$. In addition, diffraction losses in both SLMs due to the scattering and correction patterns were measured to amount to $\approx 10\%$ of additional loss. Other minor imperfections such as polarization or imaging errors might lead to slight reductions of the efficiency as well.

One reason for the difference in enhancement and efficiency between the pump and SPDC signals can be explained by material dispersion in SLM2. The phase gained by a $\lambda = 404~\mathrm{nm}$ beam is not exactly double that of a $\lambda = 808~\mathrm{nm}$, due to material dispersion, and effectively the two $808~\mathrm{nm}$ photons gain a phase that is $\approx0.7$ times the phase acquired by the pump beam. A simple simulation of this case shows that the maximal SPDC efficiency due to this dispersion effect will be $68\%$ of the pump efficiency. The remaining difference in efficiency could be explained by other experimental imperfections such as imperfect imaging between the nonlinear crystal and SLM2 or a reduction of the brightness of the SPDC source for a structured pump beam.

Two concerns that might remain regarding applicability to a real world scenario are chromatic dispersion in the atmosphere, and the fact that we use a single phase screen to emulate the effect of turbulence.

The dispersion of the atmospheric refractive index is described by $n\left(P,T,\lambda\right)=1+77.6\left(1+7.52\cdot10^{-3}\lambda^{-2}\right)\frac{P}{T}\cdot10^{-6}$, with $T$ the temperature in Kelvin, $P$ the pressure in milibars, and $\lambda$ the wavelength in $\mu\mathrm{m}$ \cite{goodman2015statistical}. It can be seen the wavelength dependence of the refractive index is indeed small, as the wavelength in $\mu m$ is on the order of a unit, and it is multiplied by $10^{-3}$. Indeed numerical simulations done in \cite{lib2020real} show the atmospheric dispersion to be negligible in the context of pump shaping. 

As for the use of a single phase screen, we note that while the statistics of scattering by atmospheric turbulence are indeed reproduced by a single phase screen, all spatial modes (or angles) passing through the emulated atmosphere will experience the same scattering, which is not the case in realistic conditions with a finite isoplanatic patch. As described in detail in\cite{lib2020real}, pump shaping can still work for thick scattering media, including atmospheric links with a finite isoplanatic patch, as long as the link's length $z$ satisfies $z<\rho_0^2/\lambda$, where $\rho_0=r_0/2.1$ is the atmospheric coherence radius. Luckily, for typical atmospheric turbulent conditions, links up to tens of kilometers can be optimized, as we show in \cite{lib2020real}. For such links, the scattering of SPDC light emitted at angles within the isoplanatic patch will be compensated, as in classical adaptive optics.

The $\rho_0$ parameter also defines how well the imaging between the crystal and SLM2 should be in our experiment, where a tolerance of a few millimeters is calculated by $\rho_0^2/\lambda$ \cite{lib2020real}. We stress that in a real life scenario these few millimeters scale up to an atmosphere with thickness of tens of kilometers, since in our experiment we scaled $r_0$ down by a factor of $1000$. 

We believe that our experimental demonstration of optimizing the spatial correlations of entangled photons through emulated turbulence is an important step towards improving the efficiency of long-range quantum links for quantum key distribution \cite{Liao2017}, especially in the case of high-dimensional entanglement and higher-order modes \cite{erhard2018twisted,erhard2020advances}.

\begin{backmatter}
\bmsection{Funding}
Zuckerman STEM Leadership Program; Israel Science Foundation (1268/16); Israeli Innovation Authority (Quantum Communication consortium).

\bmsection{Acknowledgments}
Empty. 

\bmsection{Disclosures}
The authors declare no conflicts of interest.


\bmsection{Data availability} 
Data underlying the results presented in this paper are not publicly available at this time but may be obtained from the authors upon reasonable request.

\end{backmatter}

\bibliography{ronen_references}

\end{document}